\documentclass[%
 reprint,
amsmath,
amssymb,
aps,
pra,
]{revtex4-1}

\usepackage{graphicx}
\usepackage{dcolumn}
\usepackage{bm}

\usepackage[colorlinks,
            linkcolor=blue,
            anchorcolor=blue,
            citecolor=blue]{hyperref}

\begin{document}


\title{Hybrid vector beams with non-uniform orbital angular momentum density induced by designed azimuthal polarization gradient}

\author{Lei Han, Shuxia Qi, Sheng Liu}
\email{shengliu@nwpu.edu.cn}
\author{Peng Li, Huachao Cheng}%
\author{Jianlin Zhao}%
\email{jlzhao@nwpu.edu.cn}

\affiliation{%
	MOE Key Laboratory of Material Physics and Chemistry under Extraordinary Conditions, and Shaanxi Key Laboratory of Optical Information Technology,
	School of Physical Science and Technology,
	Northwestern Polytechnical University, Xi’an 710129, China}%



\date{\today}

\begin{abstract}
	Based on angular amplitude modulation of orthogonal base vectors in common-path interference method, we propose an interesting type of hybrid vector beams with unprecedented azimuthal polarization gradient and demonstrate in experiment. Distinct to previously reported types, the synthetic hybrid vector beams exhibit geometrically intriguing projection tracks of angular polarization state on Poincare sphere, more than just conventional circles. More noteworthily, the designed azimuthal polarization gradients are found to be able to induce azimuthally non-uniform orbital angular momentum density, while generally uniform for circle-track cases, immersing in homogenous intensity background whatever base states are. Moreover, via tailoring relevant parameters, more special polarization mapping tracks can be handily achieved. These peculiar features may open alternative routes for new optical effects and applications.
	\begin{description}
		\item[Usage]
		      Secondary publications and information retrieval purposes.
		\item[PACS numbers]
		      42.60.Jf 42.25.Ja 42.15.Dp.
	\end{description}
\end{abstract}

\pacs{Valid PACS appear here}
\maketitle


\section{\label{sec:level1}INTRODUCTION}

Recent decades, vector beams have witnessed exponential increasement of research interests on itself \cite{rubinszteindunlop2017roadmap,zhan2009cylindrical}. Owing to peculiar spatially variant polarization structures, vector beams present many extraordinary optical properties, especially in tight-focusing aspect, such as optical needle and optical chain \cite{youngworth2000focusing,wang2008creation,zhao2005creation}, etc. \cite{han2018catalystlike,wang2014ultralong}. These stunning optical properties provide great potential value in various application fields, including super-resolution microscopy \cite{chen2013imaging}, optical data storage and encryption \cite{gu2014optical}, and laser fabrication \cite{cheng2017vortex-controlled}, etc. \cite{novotny2001longitudinal,xiao2018cylindrical,shang2019unidirectional,neugebauer2016polarization-controlled}. Driven by these outstanding features, constructing distinctive polarization structure has become one of topic issues \cite{li2016generation,pan2016fractal,otte2018spatial}. Up to now, diverse types of vector beams have been reported, for instance cylindrical vector beams \cite{liu2017generation}, vector beams with multiple polarization singularities \cite{han2016managing} and hybrid vector beams \cite{wang2010a,xu2016generation,lerman2010generation,beckley2010full}, etc. \cite{pan2013vector,li2018polarization,chang2017shaping}.

Hybrid vector beams are one significant and fascinating type of vector beams \cite{wang2010a,xu2016generation,lerman2010generation}. By superposition of a pair of orthogonal linear polarization basis vectors carrying spiral phase, \citeauthor{wang2010a} proposed one kind of hybrid vector beams for the first time \cite{wang2010a}. Replacing linear polarization basis states with more general elliptically polarized basis vectors \citeauthor{xu2016generation}, further produced arbitrary vector fields with hybrid polarization state \cite{xu2016generation}. In addition, Gilad generated hybrid polarized beams by transmitting radially polarized light through a wave plate \cite{lerman2010generation}, and so on. Compared with linearly polarized vector beams, besides peculiar tight focusing properties, hybrid vector beams may possess other especial features. Remarkably, azimuthally varying polarized vector beams have recently been demonstrated on its possibility of carrying orbital angular momentum (OAM) related to Pancharatnam phase \cite{yang2019manipulation,yang2017orbital-angular-momentum,zhang2015identifying,yang2016independent,pan2016arbitrarily}. Conceivably, for azimuthally varying polarized vector beams, diverse azimuthal polarization gradients may give birth to various optical properties. Hence, constructing special azimuthal polarization variation is meaningful for exploring new optical phenomena and effects. Geometrically, azimuthal polarization gradient can be intuitively indicated by projection track of angular polarization states on Poincare sphere \cite{pan2016arbitrarily}. Although various azimuthally varying polarized vector beams have been produced, up to now, most of reported cases present circular tracks on Poincare sphere \cite{zhang2015identifying,yang2016independent,pan2016arbitrarily}.

As well known, via the common-path interference method, vector beams can be generated by superposition of a pair of orthonormal polarization components carrying predesigned spatial phase structures \cite{wang2007generation,liu2012generation}. That is, the electric field distribution can be expressed as follows
\begin{eqnarray}
	{\bf{E}} = && u\left( {r,\varphi } \right)\left[ {{\alpha _0}{{\rm{e}}^{i{\delta _\alpha }}}{{\bf{e}}_\alpha } + {\beta _0}{{\rm{e}}^{i{\delta _\beta }}}{{\bf{e}}_\beta }} \right]\nonumber \\
	= && u\left( {r,\varphi } \right){{\rm{e}}^{i\frac{{{\delta _\alpha } + {\delta _\beta }}}{2}}}\left[ {{\alpha _0}{{\rm{e}}^{i\frac{{{\delta _\alpha } - {\delta _\beta }}}{2}}}{{\bf{e}}_\alpha } + {\beta _0}{{\rm{e}}^{ - i\frac{{{\delta _\alpha } - {\delta _\beta }}}{2}}}{{\bf{e}}_\beta }} \right],
	\label{eq:one}
\end{eqnarray}
where $(r,\varphi)$ is the polar coordinate, $ u(r,\varphi) $ represents the holistic complex amplitude being independent on polarization structures, $ \textbf{e}_{\alpha} $ and $ \textbf{e}_{\beta} $ are a pair of orthogonal basis vectors, $ \alpha_{0} $ and $ \beta_{0} $ denote the relative amplitude profiles of these two basis vectors with $ \left|\alpha_{0} \right|^{2} + \left|\beta_{0} \right|^{2} = 1 $, and $ \delta_{\alpha} $ and $ \delta_{\beta} $ are additional phase distributions attached by $ \textbf{e}_{\alpha} $ and $ \textbf{e}_{\beta} $, respectively. From Eq.~(\ref{eq:one}), we can know that for a preconcerted pair of basis vectors, tailoring relative amplitude and phase distributions at distinct spatial positions will give birth to diverse polarization structures. However, until now, most of reported vector beams focus on engineering relative phase distribution within spatially uniform relative amplitude profiles, i.e., the ratio of $ \alpha_{0} $ and $ \beta_{0} $ being same over entire beam cross section.

Here, by unlocking relative amplitude modulation along azimuthal coordinate, we design an intriguing kind of hybrid vector beams with unprecedented azimuthal polarization gradient, i.e., non-circle mapping track, and demonstrate few cases in experiment. The tailoring behavior of modulation parameters on polarization gradient is explored in detail. More strikingly, the designed angular polarization gradient are found to be able to induce azimuthally non-uniform OAM density within homogeneous intensity background regardless of what base states are, while generally uniform OAM density for vector beams exhibiting circular mapping tracks.

\section{\label{sec:level2}Theoretical Description}

\begin{figure}
	\centering
	\includegraphics[scale=0.4]{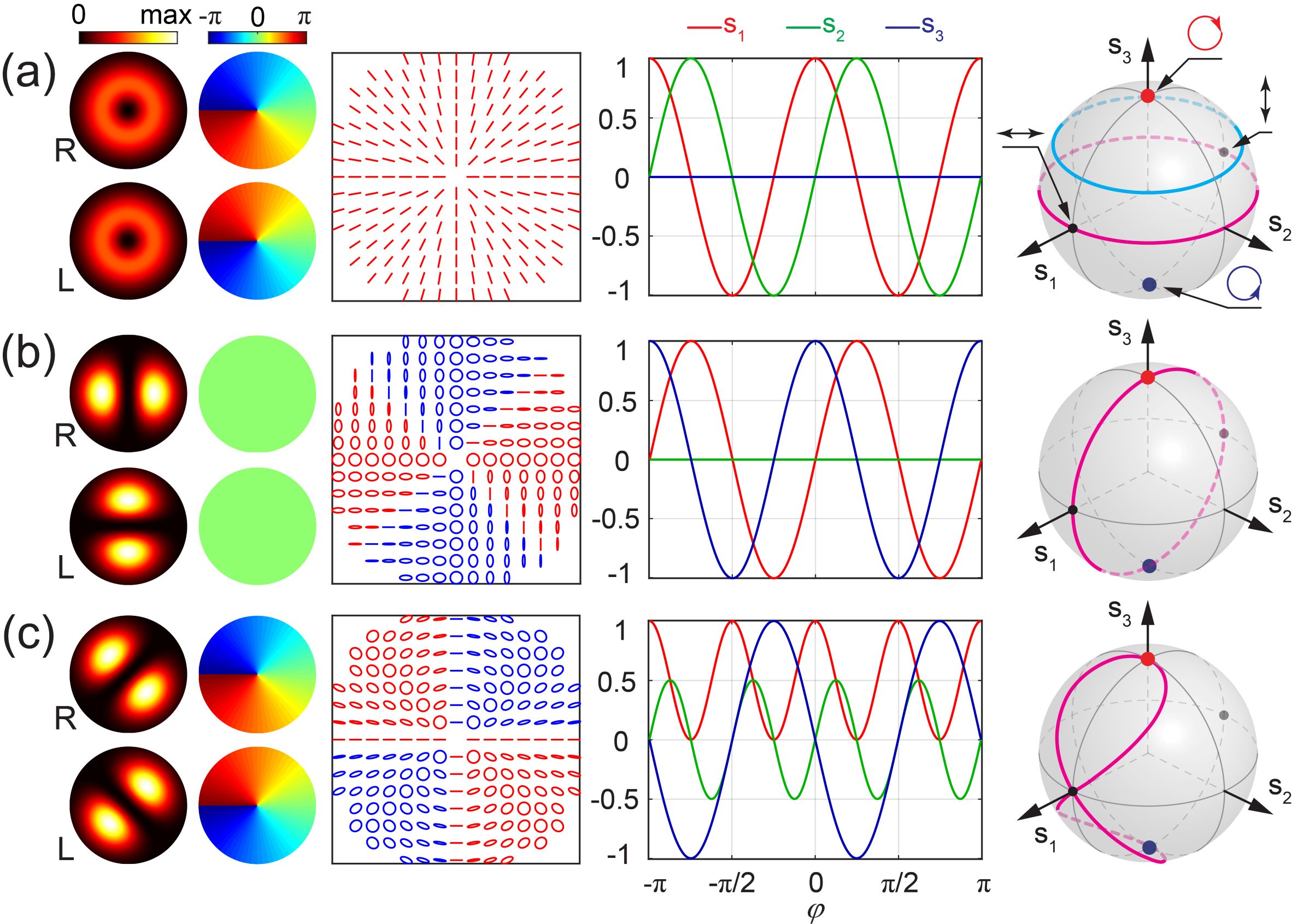}
	\caption{\label{fig:one} Illustration of the proposed scheme under circularly polarized basis vector. The cases $ (0,-1,\pi/4) $, $ (1,0,0) $ and $ (1,-1,\pi/4) $ are presented in (a)-(c) in turn. The first and second columns display the intensity and phase profiles of right- (R) and left-handed (L) circular polarization states, respectively. The synthetic polarization, normalized Stokes parameters along angular direction and polarization mapping tracks on Poincare sphere are depicted in third to fifth columns, respectively.}
\end{figure}

The proposed amplitude modulation here is described as
\begin{equation}
	{\alpha _0}\left( {r,\varphi } \right) = \cos \left( {l\varphi  + {\varphi _0}} \right), {\beta _0}\left( {r,\varphi } \right) = \sin \left( {l\varphi  + {\varphi _0}} \right),
	\label{eq:two}
\end{equation}
where $ l $ is coined as amplitude modulation factor and $ \varphi_{0} $ is initial phase with $ \varphi_{0}\in\left[0,\pi \right]  $. Considering the independence of polarization structure on $ \delta_{\alpha} + \delta_{\beta} $, hereinafter, only the cases with $ \delta_{\alpha} = -\delta_{\beta} = \textit{m}\varphi $ are discussed without losing generality. Therefore, we now get the electric field as
\begin{equation}
	{\bf{E}} = u\left( {r,\varphi } \right)\left[ {\cos \left( {l\varphi + {\varphi _0}} \right){{\rm{e}}^{im\varphi }}{{\bf{e}}_\alpha } + \sin \left( {l\varphi + {\varphi _0}} \right){{\rm{e}}^{ - im\varphi }}{{\bf{e}}_\beta }} \right].
	\label{eq:three}
\end{equation}
As seen in Eq.~(\ref{eq:three}), for any pair of basis vectors $ (\textbf{e}_{\alpha},\textbf{e}_{\beta}) $, the synthetic polarization structure is determined by parameters $ (l,m,\varphi_{0}) $. Obviously, according to whether $ l $ and $ m $ are zero, we can classify synthetic beams into the following four categories.

I. For cases of $ l = 0 $ and $ m = 0 $, the electric field will become $ {\bf{E}} = u\left( {r,\varphi } \right)\left[ {\cos {\varphi _0}{\bf{e}_\alpha} + \sin {\varphi _0}{\bf{e}_\beta}} \right] $. It can be clearly known that the generated light beams are scalar beams for arbitrary basis vectors $ (\textbf{e}_{\alpha},\textbf{e}_{\beta}) $. And the exact polarization state is determined by the value of $ \varphi_{0} $.

II. For cases of $ l = 0 $ while $ m \neq 0 $, if the initial angle $ \varphi_{0} = 0, \pi/2, $ or $ \pi $, only one base state has non-zero complex amplitude, hence, the resulting beam is a scalar beam carrying spiral phase. While for cases with $ \varphi_{0} \neq 0, \pi/2, $ and $ \pi $, conventional azimuthally varying polarized vector beams are achieved as presented in Refs.~\onlinecite{zhang2015identifying,yang2016independent,pan2016arbitrarily}. In other words, the previously reported cases can actually be viewed as special cases of our general description. For these vector beams, the polarization mapping tracks on Poincare sphere display circular shape. The plane of circular track is perpendicular to the connection line of the base vectors. And the central position and diameter of circles are dependent on base vectors and $ \varphi_{0} $  \cite{yang2016independent}.

To illustrate and manifest corresponding properties of vector beams intuitively, we present one case of $ (0,-1,\pi/4) $ in Fig.~\ref{fig:one}(a). Without losing generality, here we take circularly polarized base vectors as representative, i.e., $ (\textbf{e}_{\alpha},\textbf{e}_{\beta}) = (\textbf{e}_{R},\textbf{e}_{L}) $. As shown, the right- and left-hand polarized components have same intensity profile, while opposite spiral phase maps. Hence, the generated beam is well-known radial vector beam. The synthetic polarization structure is depicted in third column, where red and blue ellipses represent polarization ellipse for $ S_{3} \geq 0 $ and $ S_{3} \leq 0 $, respectively. In calculations, we set the global amplitude $ u(r,\varphi) $ as 1st order Bessel Gaussian profile, i.e., $ u(r,\varphi) = \exp (-r^{2}/w_{0}^{2})J_{1}(2r/w_{0}) $, where $ w_{0} = 9 mm $ is the beam waist. Moreover, the evolution curve of normalized Stokes parameters is plotted in fourth column to show polarization variation along angular direction. Further, we draw the polarization mapping track on Poincare sphere in pink color in fifth column. Meanwhile, we also present the mapping track of another case $ (0,-1,\pi/6) $ in cyan color. Visually, as mentioned above, various values of $ \varphi_{0} $ give different radiuses. And both circle planes are perpendicular to the linking line of right- and left-handed circular polarization states.

III. For $ l \neq 0 $ while $ m = 0 $, the electric field is $ \textbf{E} = u(r,\varphi)\left[ \cos \left( {l\varphi + \varphi_{0}} \right) \textbf{e}_{\alpha} + \sin \left( {l\varphi + \varphi_{0}} \right) \textbf{e}_{\beta}\right] $. Therefore, the synthetic light beams are also azimuthally varying polarized vector beams. Geometrically, the polarization mapping tracks on Poincare sphere are always meridian circles. However, distinct to the cases in II, the circular track will travel through the two base vectors. As seen from case $ (1,0,0) $ depicted in Fig.~\ref{fig:one}(b), a non-zero $ l $ gives non-uniform relative amplitude ratio of two orthogonal components. Thus, a hybrid vector beam having meridian mapping circle is constructed.

IV. More noteworthily, when both $ l $ and $ m $ are non-zero, i.e., $ l \neq 0 $ and $ m \neq 0 $, an unusual kind of vector beams will be constructed, which is the most intriguing core of the proposed scheme. As a representative, we give the case $ (1,-1,\pi/4) $ in Fig.~\ref{fig:one}(c). Obviously, significantly different from circle projection tracks in Figs.~\ref{fig:one}(a) and \ref{fig:one}(b), there is a distinctive mapping track seem like an infinity symbol in Fig.~\ref{fig:one}(c). The special mapping tracks indicate straightly that the proposed scheme can enable different polarization states on Poincare sphere to establish unprecedented connection, thus possess capability to configurate intriguing azimuthal polarization gradient.

\begin{figure}
	\centering
	\includegraphics[scale=0.35]{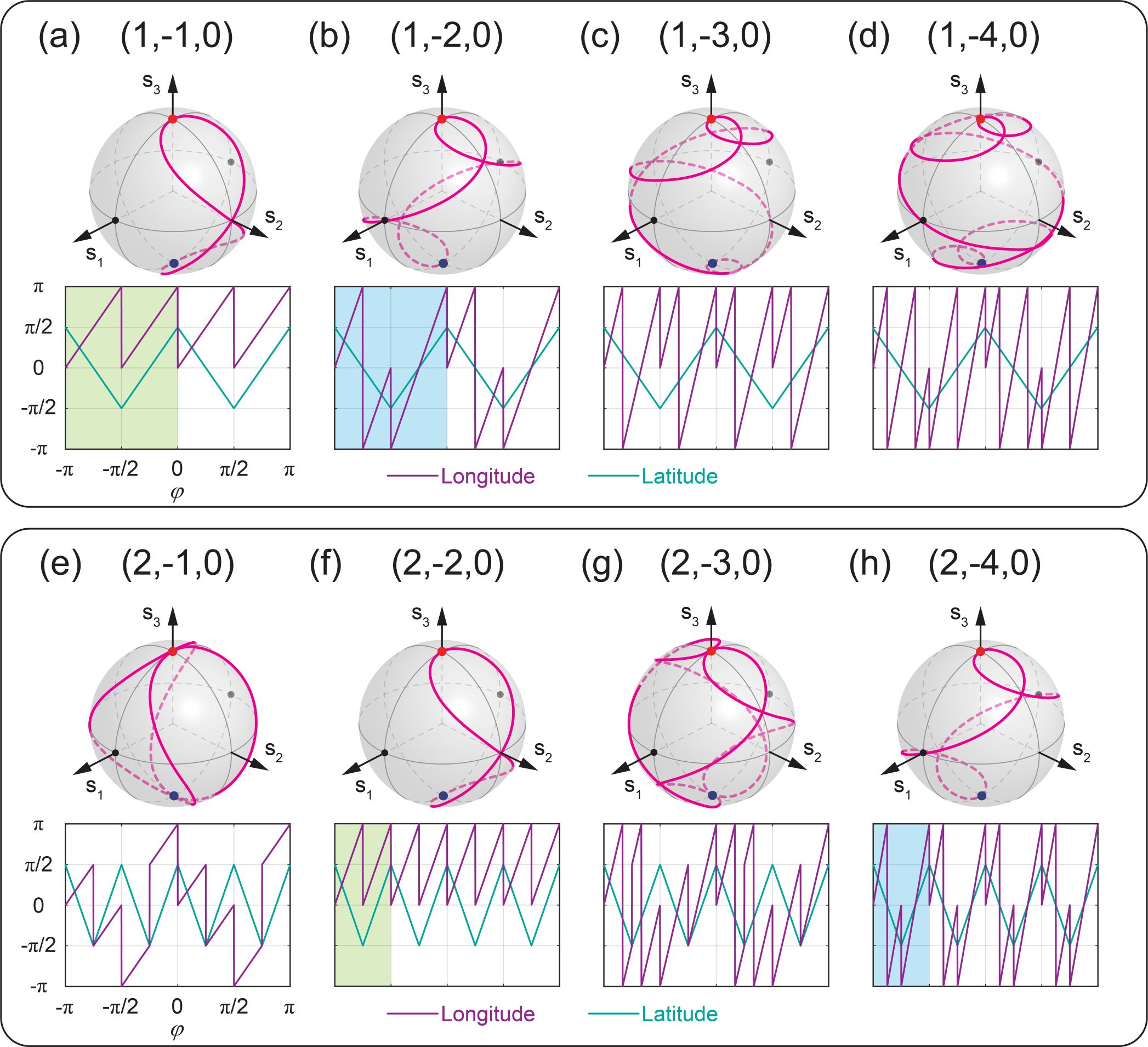}
	\caption{\label{fig:two} Modulation effect of $ l $ and $ m $ on polarization gradient. The polarization structures, the polarization mapping tracks and the evolution curves of longitude and latitude angles along angular direction are depicted in first to third rows, respectively.}
\end{figure}

To reveal the modulation effect of $ l $ and $ m $ on azimuthal polarization gradient, in Figs.~\ref{fig:two}(a)-\ref{fig:two}(h), we further depict some cases with different $ l $ and $ m $. As well known, besides Stokes parameters, any polarization state can be briefly described by a set of angle coordinates $ (2\psi,2\chi) $ on Poincare sphere \cite{xu2016generation,zhang2015identifying}. Here, $ 2\psi $ and $ 2\chi $ stand for longitude and latitude angles of polarization state, respectively. In second row of Fig.~\ref{fig:two}, we plot the varying curve of longitude and latitude along azimuthal coordinate. According to the relationship of Stokes parameters and longitude and latitude angles, we can get the corresponding result as follows
\begin{equation}
	\tan 2\psi = - \tan (2m\varphi),\sin 2\chi = \cos \left( {2l\varphi + 2{\varphi _0}} \right).
	\label{eq:four}
\end{equation}

Comparing the results in Figs.~\ref{fig:two}(a) with \ref{fig:two}(f) and in Figs.~\ref{fig:two}(b) with \ref{fig:two}(h), and together with more other cases, it is found that for different sets of $ (l,m) $ with same ratio of $ l/m $, the polarization mapping tracks will share same shape. And various ratios will give rise to diverse interesting shaped tracks. Furthermore, via analyzing variation of longitude and latitude angles along angular coordinate, we find that for certain ratio of $ l/m $, the specific values of $ l $ and $ m $ decide the polarization variation period along azimuthal coordinate, as displayed in Figs.~\ref{fig:two}(a) and \ref{fig:two}(f) as well as Figs.~\ref{fig:two}(b) and \ref{fig:two}(h). Moreover, we can get the fact that $ l $ determines the gradient of latitude, and $ m $ together with $ l $ controls the gradient of longitude. Here, it should be noted that though the tangent value of longitude angle is only determined by $ m $, owing to periodical feature of tangent function, the longitude angle will have distinct varying curve associated with $ l $, such as seen in Figs.~\ref{fig:two}(a) and \ref{fig:two}(e). In addition, comparing the result in Fig.~\ref{fig:one}(c) with that in Fig.~\ref{fig:two}(a), it is found that $ \varphi_{0} $ plays a role of controlling position of mapping track on Poincare sphere. All above mentioned features can be verified by more other cases.

\begin{figure}
	\centering
	\includegraphics[scale=0.4]{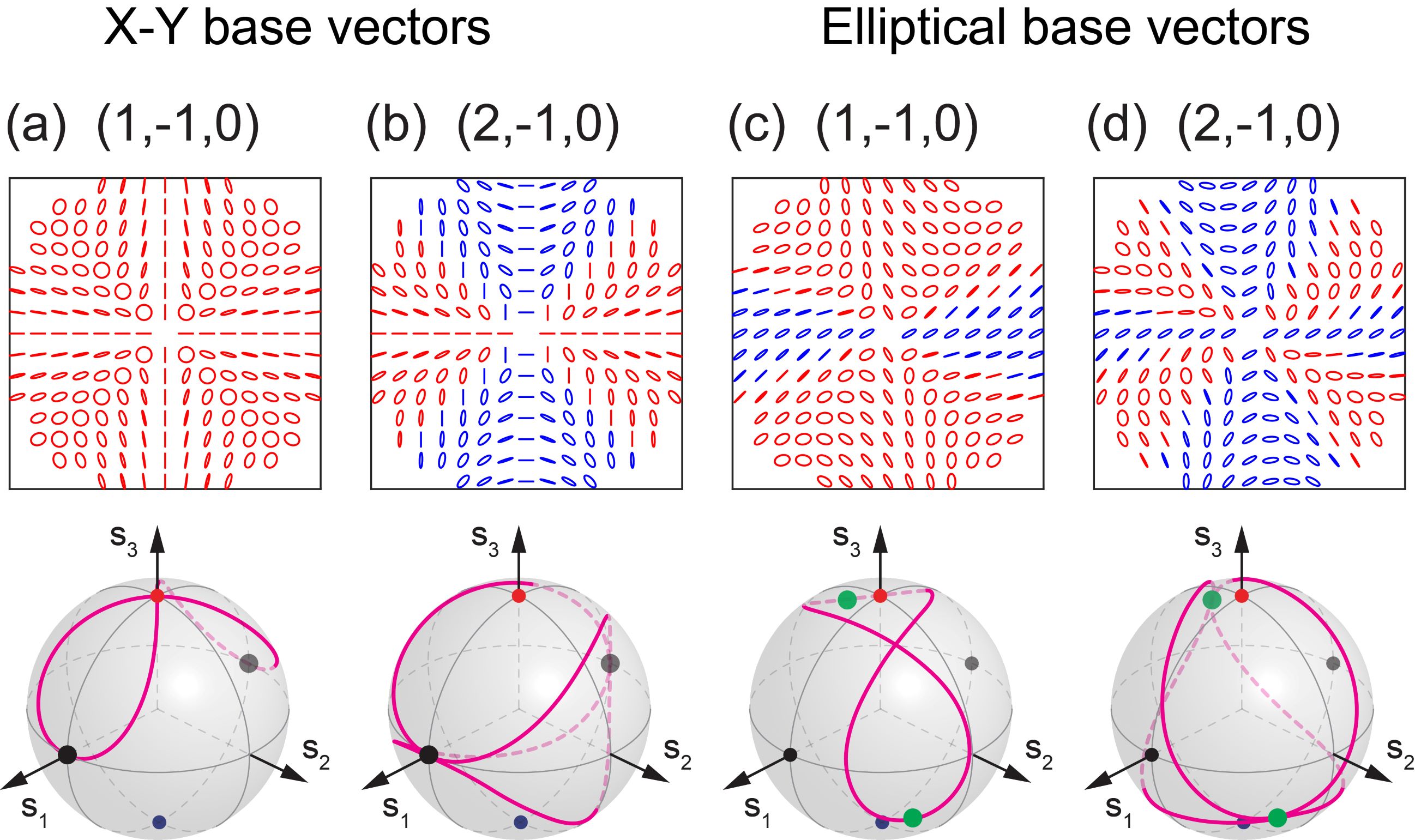}
	\caption{\label{fig:three} Hybrid vector beams under various base states.}
\end{figure}

In fact, the aformentioned tailoring behavior of $ l $ and $ m $ on polarization gradient applies to any orthogonal base vectors. As shown in Fig.~\ref{fig:three}, it can be clearly seen that under different base vectors, when hybrid vector beams have same $ l/m $, the polarization mapping tracks will share same shape. And the position of mapping track on Poincare sphere depends on the base vectors. Moreover, if we set arbitrary $ \textbf{e}_{\alpha} $ and $ \textbf{e}_{\beta} $ as north and south poles respectively, the modulation feature of $ l $ and $ m $ on latitude and longitude angles is same as that under circularly polarized base vectors. In Figs.~\ref{fig:three}(c) and \ref{fig:three}(d), the green points on Poincare sphere stand for the chosen elliptical polarized base states, and their longitude and latitude angles are $ (\pi/3,-\pi/6) $ and $ (-2\pi/3,\pi/6) $, respectively.

\section{\label{sec:level3}Non-uniform OAM density associated with azimuthal polarization gradient}

By analyzing OAM density of hybrid vector beams with $ l \neq 0 $ and $ m \neq 0 $ in theory, we excitingly find that the proposed azimuthal polarization gradient can induce fancy non-uniform OAM density. As presented in Ref.~\onlinecite{pan2016arbitrarily}, for azimuthally varying polarized vector beams, regardless of $ \textbf{e}_{\alpha} $ and $ \textbf{e}_{\beta} $, the $ z $ component of OAM density comprises two parts associated with azimuthal gradient, that is,
\begin{equation}
	{l_z} = {\left( {{\bf{r}} \times {\bf{p}}_ \bot ^o} \right)_z} \propto {u^2}\frac{{\partial \psi }}{{\partial \varphi }} + {u^2}{\mathop{\rm Im}\nolimits} \left[ {{\alpha ^ * }\frac{{\partial \alpha }}{{\partial \varphi }} + {\beta ^ * }\frac{{\partial \beta }}{{\partial \varphi }}} \right],
	\label{eq:five}
\end{equation}
where $ \psi $ represents the phase distribution and is corresponding to $ (\delta_{\alpha} + \delta_{\beta})/2 $ in our discussed cases. Besides, there is
\begin{equation}
	\alpha  = \cos \left( {l\varphi  + {\varphi _0}} \right){{\rm{e}}^{im\varphi }},
	\beta  = \sin \left( {l\varphi  + {\varphi _0}} \right){{\rm{e}}^{ - im\varphi }}.
	\label{eq:six}
\end{equation}
As shown in Eq.~(\ref{eq:five}), the first term is induced by phase gradient like conventional scalar vortex phase. In contrast, the second term in Eq.~(\ref{eq:five}) is associated with polarization gradient in angular direction. Substituting Eq.~(\ref{eq:six}) into Eq.~(\ref{eq:five}), we can have polarization-gradient-related OAM density as follows
\begin{equation}
	{l_{z2}} \propto 2m\cos \left( {2l\varphi  + 2{\varphi _0}} \right){u^2}.
	\label{eq:seven}
\end{equation}
Remarkably, for azimuthally uniform amplitude distribution, i.e., $ u(r,\varphi)=u(r) $, the polarization-gradient-related OAM density is closely relevant to $ (l,m,\varphi_{0}) $. Obviously, for cases with $ l = 0 $, the OAM density profile is spatially uniform, while for cases with $ l \neq 0 $, the OAM density features variation along azimuthal coordinate $ \varphi $. Thus, a non-zero $ l $ together with a non-zero $ m $ will give rise to the non-uniform OAM density. And $ m $ plays a role on controlling the maximum amplitude of local OAM density. That is, we can achieve azimuthally variant OAM density under homogeneous intensity background, which is in fact rare for conventional scalar and vector beams with azimuthally varying polarization states as reported previously \cite{zhang2015identifying,yang2016independent,pan2016arbitrarily}. Moreover, holding the same intensity background, diverse OAM density distributions can be achieved via tailoring aforementioned modulation parameters.

\begin{figure}
	\centering
	\includegraphics[scale=0.4]{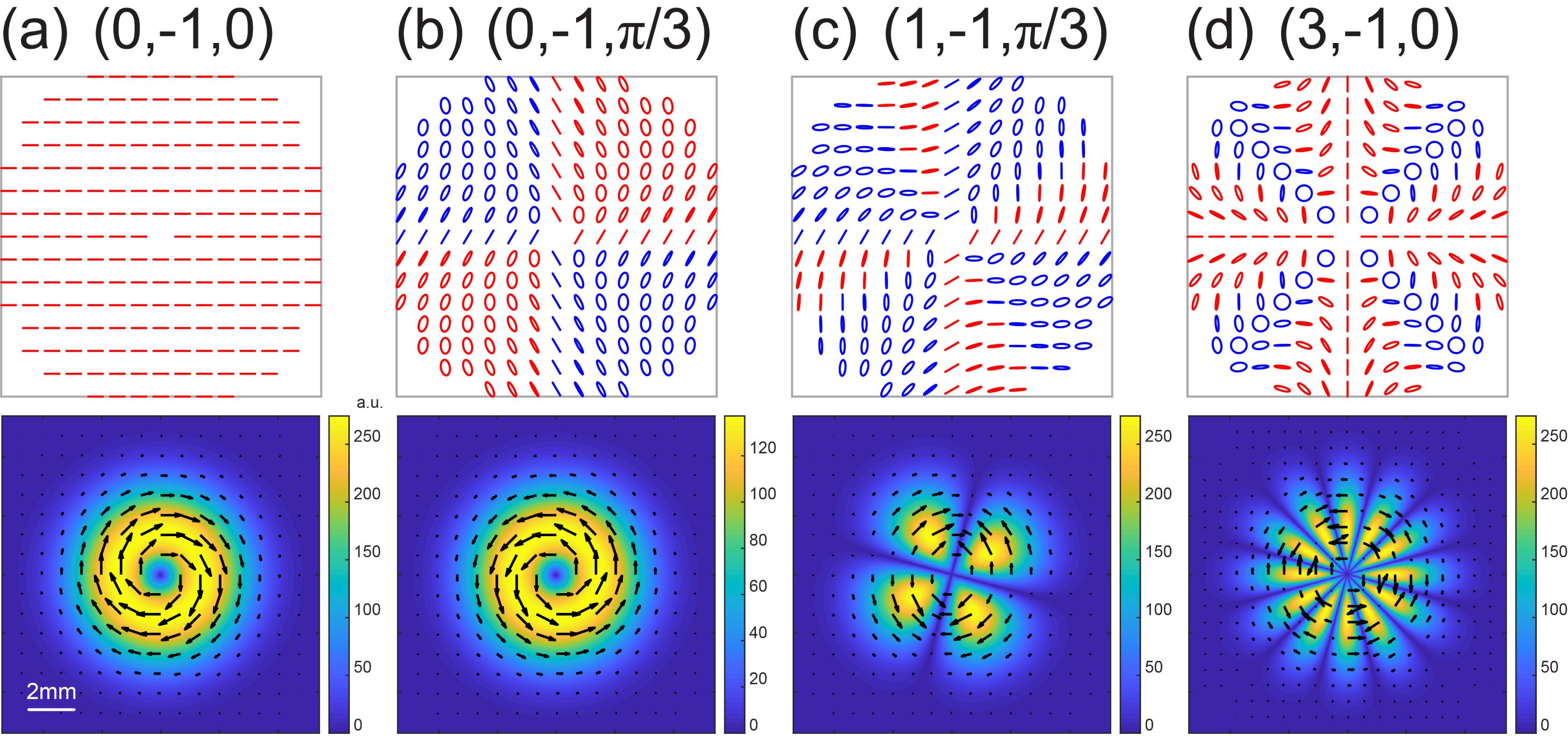}
	\caption{\label{fig:four} (a)-(d) Polarization structures (upper) and transverse Poynting vector patterns (lower) of hybrid vector beams with $ (0,-1,0) $, $ (0,-1,\pi/3) $, $ (1,-1,\pi/3) $, $ (3,-1,0) $ under X-Y base vector.}
\end{figure}

For illustration and considering the proportional relation of linear momentum density with Poynting vector, in Fig.~\ref{fig:four}, we show the transverse part of Poynting vector, i.e., transverse energy flux density distribution, of few cases under X-Y base vector. The Poynting vector distribution can indirectly reflect the polarization-gradient-related OAM density. As we can see, in Fig.~\ref{fig:four}(a), owing to $ l = 0 $ and $ \varphi_{0} = 0 $, the vector beam degenerates into a $ x $-polarized scalar beam, therefore, the azimuthally uniform transverse Poynting vector totally originates from spiral phase, where black arrows indicate the direction. For case with $ \varphi_{0} = \pi/3 $, a hybrid vector beam with circular mapping track is constructed as shown in Fig.~\ref{fig:four}(b). Hence, the transverse Poynting vector is induced by azimuthally varying polarization structures and keeps cylindrical symmetry. Nevertheless, notably, the maximum value becomes lower than that in Fig.~\ref{fig:four}(a), which is in consistent with the result in Ref.~\onlinecite{pan2016arbitrarily}. In contrast to Figs.~\ref{fig:four}(a) and \ref{fig:four}(b), in agreement with predictions according to Eq.~(\ref{eq:seven}), a non-zero $ l = 1 $ in Fig.~\ref{fig:four}(c) and $ l = 3 $ in Fig.~\ref{fig:four}(d) induce periodically inhomogeneous transverse Poynting vector along azimuthal direction. And as expected, the variation period can be manipulated via adjusting $ l $. The unique azimuthally inhomogeneous OAM density may draw different phenomena in optical force and manipulation.

\section{\label{sec:level4}Experimental setup and results}

\begin{figure}
	\centering
	\includegraphics[scale=0.85]{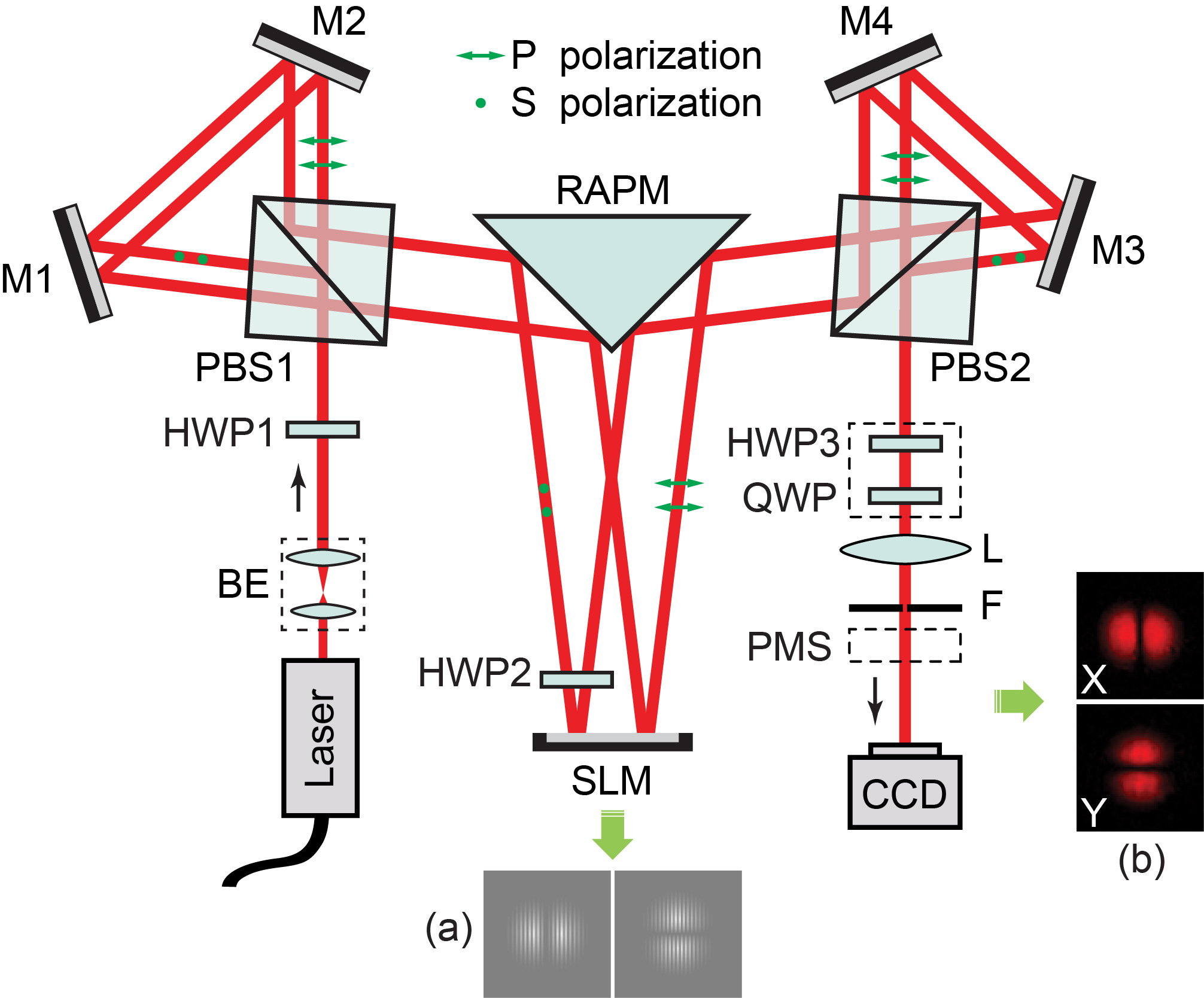}
	\caption{\label{fig:five} Experimental setup. BE, beam expansion module; HWP, half-wave plate; PBS, polarizing beam splitter; M1-M4, mirrors; RAPM, right-angle prism mirror; SLM, spatial light modulator; QWP, quarter-wave plate; L, lens; F, spatial filter; PMS, polarization measuring system; CCD, charge coupled device. Insets: (a) The computer-generated holograms and (b) the measured intensity patterns of $ x $ and $ y $ components for the case $ (1,0,0) $ on the base of X-Y basis vectors, i.e., radial vector beam.}
\end{figure}

To realize the proposed hybrid vector beams in experiment, it is crucial to steer the relative amplitude profiles of two orthogonal constituents. Considering this condition, we employ the efficient experimental setup proposed by \citeauthor{liu2018highly} \cite{liu2018highly}, as shown in Fig.~\ref{fig:five}. For detailed description on principle, please refer to Section 2 in Ref.~\onlinecite{liu2018highly}. Note, in our experiment, a half-wave plate (HWP3) and a quarter-wave plate (QWP) are needed to transform the synthetic light beam into aimed vector beam. To this end, the angles between $ x $ direction and fast axes of HWP3 and QWP need to be adjusted properly in order to alter $ p $ and $ s $ components into arbitrary prescribed basis vectors. The produced polarization structures are measured via a polarization measuring system (PMS) proposed in Ref.~\onlinecite{liu2017a}. Using the encoding technique pioneered in Ref.~\onlinecite{davis1999encoding}, both amplitude and phase information for any $ l $, $ m $ and $ \varphi_{0} $ can be conveniently encoded onto one phase map displayed in SLM.

\begin{figure}
	\centering
	\includegraphics[scale=0.4]{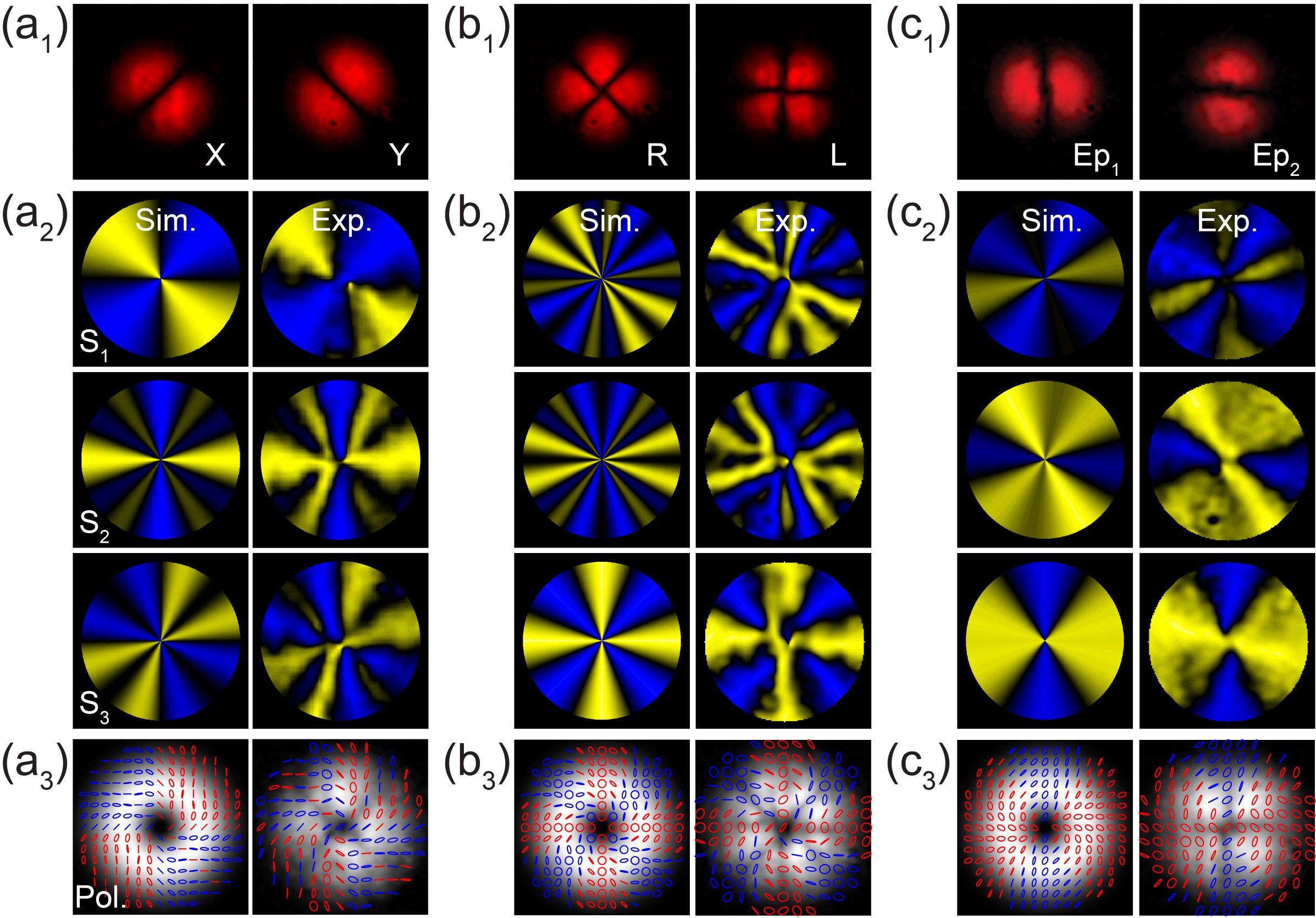}
	\caption{\label{fig:six} Experimental and theoretical results for (a) $ (1,-2,\pi/4) $ under X-Y basis vector, (b) $ (2,-3,0) $ under R-L basis vector and (c) $ (1,-1,0) $ under elliptical basis vector Ep$ _{1} $-Ep$ _{2} $. The longitude and latitude of Ep$ _{1} $ and Ep$ _{2} $ on Poincare sphere are $ (-\pi/3,\pi/3) $ and $ (2\pi/3,-\pi/3) $, respectively. The measured intensity patterns of orthogonal basis vectors are shown in (a$ _{1} $)-(c$ _{1} $). Theoretical (left) and experimental (right) results of Stokes parameters and produced polarization structures are depicted in (a$ _{2} $)-(c$ _{2} $) and (a$ _{3} $)-(c$ _{3} $) in turn.}
\end{figure}

Figures~\ref{fig:six}(a)-\ref{fig:six}(c) exhibit the theoretical and experimental results of three representative cases under X-Y, R-L and a pair of elliptical basis vectors Ep$ _{1} $-Ep$ _{2} $, respectively. The longitude and latitude of the chosen elliptical basis vectors Ep$ _{1} $-Ep$ _{2} $ on Poincare sphere are $ (-\pi/3,\pi/3) $ and $ (2\pi/3,-\pi/3) $, respectively. The corresponding modulation parameters for each case are in turn $ (1,-2,\pi/4) $, $ (2,-3,0) $ and $ (1,-1,0) $. As shown, evidently, the generated polarization structure as well as measured Stokes parameters are in good agreement with the theoretical results, which provides a proof of feasibility of the proposed scheme. The differences between experimental and theoretical results may owe to many factors, including quality of laser beam, imaging quality of CCD and post-processing program, etc.

\section{\label{sec:level5}CONCLUSIONS}
In conclusion, based on angular amplitude modulation of orthogonal base states, we propose and demonstrate an unusual type of hybrid vector beams with intriguing azimuthal polarization gradient. Geometrically, the synthetic polarization structures here present unprecedented non-circle mapping tracks on Poincare sphere. More interestingly, the produced peculiar azimuthal polarization gradient are found to able to induce azimuthal non-uniform OAM density within homogenous intensity background, while in general uniform for circle-tracks cases. Moreover, the polarization gradient, i.e., polarization mapping tracks on Poincare sphere, can be simply and flexibly manipulated via arranging relevant modulation parameters. These features may provide a new way for focal fields shaping and local OAM tailoring etc.

\begin{acknowledgments}
	This work was supported by National Key Research and Development Program of China (2017YFA0303800), National Science Foundation of China (NSFC) (11634010, 61675168, 91850118, 11774289, 11804277), Fundamental Research Funds for the Central Universities (3102019JC008), and Basic research plan of natural science in Shaanxi province (2018JM1057, 2019JM-583).
\end{acknowledgments}


\nocite{*}
\bibliography{mybib}

\end{document}